# Colossal Anomalous Hall Conductivity and Topological Hall Effect in Ferromagnetic Kagome Metal Nd$_3$Al

Durgesh Singh[1], Jadupati Nag[1], Sankararao Yadam[2], V. Ganesan[3], Aftab Alam[1] and K.G. Suresh[1*]

**Abstract**

Historically, the genesis of anomalous Hall effect (AHE) in magnetic materials has always been a fascinating yet controversial topic in the solid state physics community. Recent progress on the understanding of this topic has revealed an intimate connection between the Berry curvature of occupied electronic states and the intrinsic AHE. Magnetic Weyl semimetals with broken time reversal symmetry is a classic example, which is expected to show large contributions to Berry curvature around the topological nodes and hence to the AHE. Here, we report a kagome metallic ferromagnet Nd$_3$Al, with a large unconventional positive magnetoresistance (~ 80 %) and colossal anomalous Hall conductivity of $1.8 \times 10^5$ S/cm (largest ever reported to the best of our knowledge). We also show that the magnetic state of this compound is quite different from its analogues in many respects. While the compound is predominantly an itinerant ferromagnet, its low temperature phase exhibits topological band structure, enhanced skew scattering as well as topological spin texture arising in the spin frustrated kagome lattice. Various experimental findings such as topological Hall effect, non-saturating positive magnetoresistance etc. give strong indication to this scenario. Ab-initio calculations broadly confirm the experimental findings by revealing the presence of flat bands and Weyl points originating from the itinerant Nd moments. The non-trivial band structure, enhanced skew scattering and the spin texture in a clean polycrystalline sample are found be responsible for the colossal Hall conductivity and topological Hall effect.

Affiliations

[1]Department of Physics, Indian Institute of Technology Bombay, Mumbai 400076, Maharashtra, India

[2]Humanities and Sciences Department, CVR College of Engineering, Rangareddy, Hyderabad 501510, Telangana, India

[3]Medi-Caps University, A.B. Road, Indore 453331, Madhya Pradesh, India

*Corresponding author - Email: suresh@phy.iitb.ac.in

# INTRODUCTION

Along with the discovery of novel topological insulators, recently topological semimetals (TSMs) have drawn tremendous research interest in condensed matter physics. TSMs are known to be topologically robust and are classified into several types such as Dirac, Weyl, triple-point, double Weyl semimetals etc. based on the degeneracy of the band-crossing points (1, 2). The non-trivial band topology gives rise to exotic properties like anomalous transport in TSMs (3-7). Dirac semimetals (DSMs) have symmetry protected band crossings near the Fermi level ($E_F$), while Weyl semimetals (WSMs) are characterized by Weyl nodes near $E_F$ due to the broken time reversal symmetry (TRS) arising from their magnetic order. This gives rise to large anomalous Hall effect (AHE) originating from the significant contribution to Berry curvature around the nodes (8-11). In addition to this intrinsic contribution, extrinsic mechanisms based on spin-orbit coupling also plays a significant role in enhancing the AHE.

If the topological character is associated with magnetic frustration, then it exhibits Dirac and chiral quantum spin liquid (QSL) nature in magnetic fields, as observed in α-$RuCl_3$ (12). $Co_3Sn_2S_2$ is a classic example of Weyl semimetal which shows large AHE due to the coexistence of ferromagnetism (FM) and geometric frustration resulting in spin glass behavior and exchange bias (13). Theoretical results on heavy fermion Kondo alloy $Ce_3Al$ perceive kagome-derived flat bands and Dirac cones in its quasiparticle band structure, which is predicted as the first member of heavy-fermion kagome system (14). In fact, kagome lattice is being seen as an important platform to host novel and interesting topological phases. Though existence of Dirac nodes in such lattices is often seen, appearance of Weyl nodes in magnetic kagome lattice is rather rare. These observations motivated us to study $Nd_3Al$, which belongs to the $RE_3Al$ (RE= rare earth) family (15-17). $Nd_3Al$ crystallizes in the prototype hexagonal $Ni_3Sn$ structure (space group $P6_3/mmc$)(18). There exist some papers on this system, describing the usual transport and magnetic properties. Fukuhara et al. (19) reported the specific heat coefficient (γ) and the Debye temperature ($\theta_D$) to be 42 mJ/mol-$K^2$ and 170 K respectively. It has an anisotropic magnon excitation gap of ~ 30 K and is expected to be a localized ferromagnet. The dc susceptibility studies revealed long-range magnetic order below 74 K. The electronic conduction mechanism is confirmed by negative Hall coefficient ($R_O$). Unlike $Ce_3Al$, it neither exhibits Kondo effect nor undergoes a structural transition from hexagonal to low temperature monoclinic structure (16,20). Furthermore, a large positive magnetoresistance (LPMR) and Hall conductivity below 20 K and a high value of γ rule out a simple FM ground state. In this communication, we report a colossal anomalous Hall conductivity (AHC~$1.8 \times 10^5$ S/cm) for $Nd_3Al$, which, to the best of our knowledge, is the largest ever reported. It also indirectly hints at the presence of topological Hall effect in a polycrystalline sample with a very high residual resistivity ratio (RRR) of 83.2. We also found an unconventional LPMR (~ 80%) with a strikingly different magnetic state as compared to other compounds belonging to this family. Though $Nd_3Al$ is predominantly an itinerant ferromagnet, its low temperature phase exhibit topological band structure, as confirmed by our first principles calculations. Our ab-initio calculations not only demonstrate the origin of giant AHC associated with the non-trivial band topology in $Nd_3Al$, but also reveal the existence of Weyl nodes near $E_F$ and hence the associated Berry curvature contribution to the AHE.

## RESULTS AND DISCUSSION

### Structural Analysis

XRD data was refined with the Full-Profile Rietveld method using the program FullProf Suite (21). The pattern (Fig. 1) confirms that $Nd_3Al$ crystalizes in the hexagonal structure with single phase with lattice parameters $a = b = 7.00$ Å and $c = 5.43$ Å. The GoF-index is 1.2 and $\chi^2$ is 1.41, which confirm the goodness of the fit. A very weak, unidentified peak is seen at 31.46°, which remained in all batches, similar to the one reported in ref. (22).

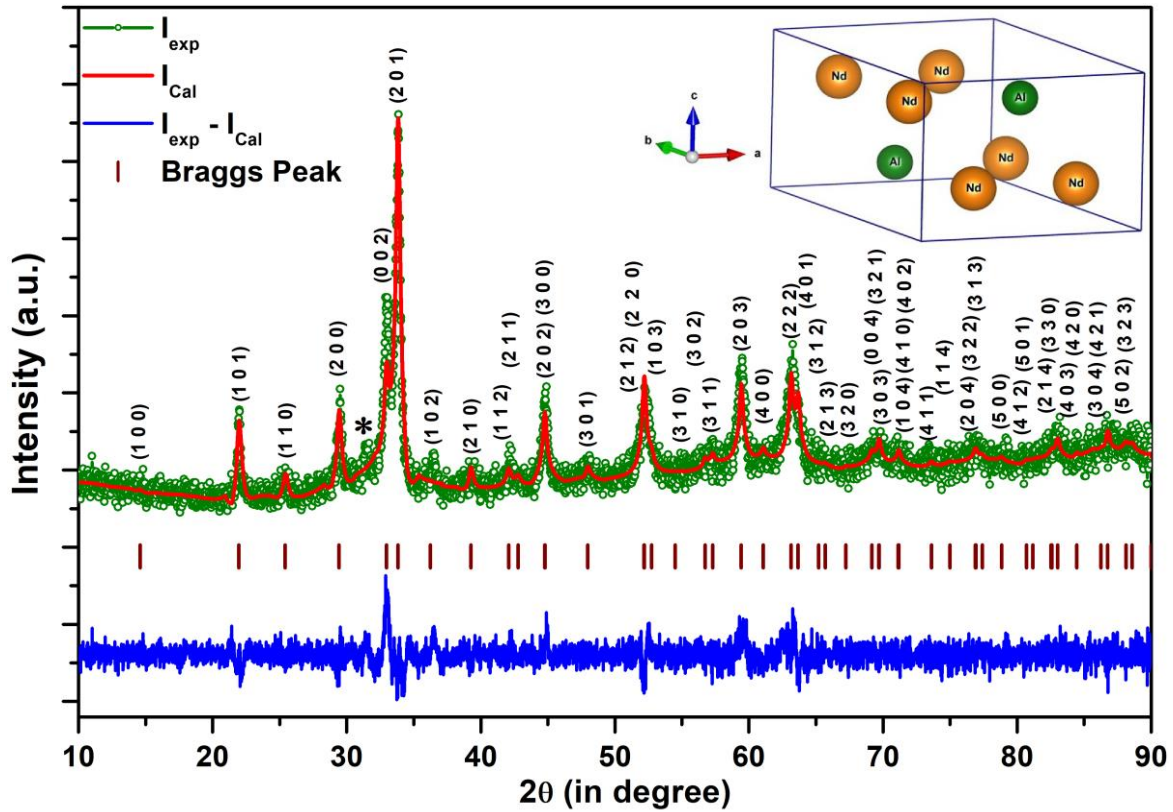

**Figure 1.** XRD pattern of $Nd_3Al$ recorded at room temperature (dark Green). Solid line through the experimental point is the Rietveld refined profile for the $Ni_3Sn$ – type hexagonal structure (Red). The short vertical bars mark the Bragg peak

positions. The lowermost curve represents the difference between the experimental and the calculated intensities. Inset shows the crystal structure of Nd$_3$Al. Here, Nd and Al atoms are represented by orange and green spheres, respectively.

## Magnetization

To investigate the magnetic behaviour of the sample, M (T) has been measured in different fields and shown in fig. 2 (a). DC susceptibility χ (T) (left scale) in zero field cooling (ZFC), field cooled cooling (FCC) and field cooled warming (FCW) protocols, and the inverse susceptibility (right scale) in the FCW mode are shown in figure 2 (b) in the range of 2 - 300 K in 0.1 T. The dip in dχ/dT vs. T (inset of figure 2 (b)) corresponds to $T_C$ ~ 74 K, which agrees with the reported value [17]. There is a bifurcation between ZFC and FCC/ FCW data below $T_C$, which indicates the magnetocrystalline anisotropy (especially due to Nd$^{3+}$) or magnetic frustration. Fit of the FCW data to χ (T) = $χ_0$ + C/(T-$θp$), [$χ_0$ is the temperature independent susceptibility, C is the Curie constant and $θ_P$ is the Weiss temperature] has yielded the effective moment, $μ_{eff}$ = 5.47 $μ_B$/f.u. or 1.82 $μ_B$ / Nd$^{3+}$ and $θp$ = 74.5 K. The high value of $χ_0$ (~10$^{-3}$ emu/mol – Oe) obtained can be due factors such as heavy fermion nature, magnetic disorder or spin/valence fluctuations (i.e., itinerant nature of Nd$^{3+}$ moments), whereas the positive value of $θp$ indicates the predominant FM interaction. The estimated $μ_{eff}$ is less than the free Nd$^{3+}$ moment (3.62 $μ_B$) as well as the theoretically calculated value (See TABLE I). The deviation of $χ^{-1}$ (T) vs. T curve from the Curie-Weiss fit and the reduction in the $μ_{eff}$ can be attributable to crystal field effect and/or disorder in the geometrically frustrated lattice, as other factors like Kondo effect and mixed valency are rarely observed in Nd compounds.

In fig. 2 (c), It is observed that M-H curves below $T_C$ show a sharp increase like in a FM and the magnetization nearly saturates. The sample appears to be a soft FM, with a small hysteresis below 30 K [as shown in the inset of figure 2 (c)]. Interestingly, the 9 T magnetization at 20 K is found to be higher than that in the range 2 - 10 K, which indicates a complex magnetic structure below 20 K, along with Nd$^{3+}$ ions either having partially lost their moments or undergone magnetic frustration. All these observations reveal the coexistence of a major FM order along with some degree of frustration or a different spin texture, at low temperatures.

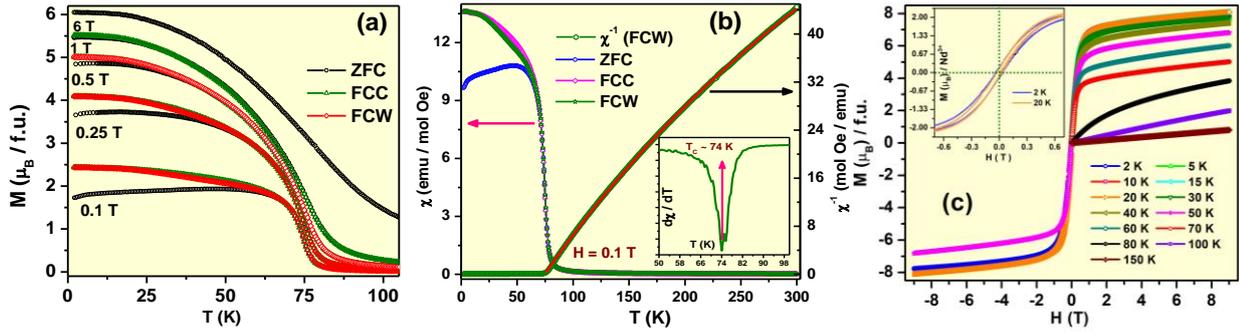

**Figure 2.** **(a)** M vs. T curves in ZFC and FCC/FCW protocol up to 1 T and ZFC curve in 6 T. Above 1 T, bifurcation between ZFC and FCC/FCW vanishes. **(b)** dc magnetic susceptibility ($\chi$) [left scale] in ZFC, FCC & FCW protocols and FCW inverse susceptibility [Right scale] ($\chi^{-1}$) vs. T of $Nd_3Al$. $T_C$ obtained from the $d\chi/dT$ in FCW mode is shown in the inset. FCW inverse dc susceptibility ($\chi^{-1}$) curve is fitted with modified Curie Weiss law as shown by red line. **(c)** Magnetic field dependence of dc magnetization data, M(H) measured between 2 and 150 K in fields up to 9 T for $Nd_3Al$. Inset shows the M(H) curves for 2 and 20 K.

## AC Susceptibility:

Figure 3 shows the real ($\chi'$) and the imaginary ($\chi''$) components of the AC susceptibility (ACS) data. Magnetic transition seen in the ACS data coincides with the Curie temperature obtained from the $\chi_{dc}$ – T plot. However, one can see in the real part of ACS that there is a sharp change at $T_C$ and below it the data decreases slowly with a slope change at T = 35 K [fig. 3(a)]. Therefore, we can conclude that the moments relax with different relaxation rate below $T_C$. Moreover, the data show a positive shift with frequency in this temperature regime, as observed in canonical or re-entrant spin glasses. The absence of a well-defined peak at $T_C$ is seen in the heat capacity data as well.

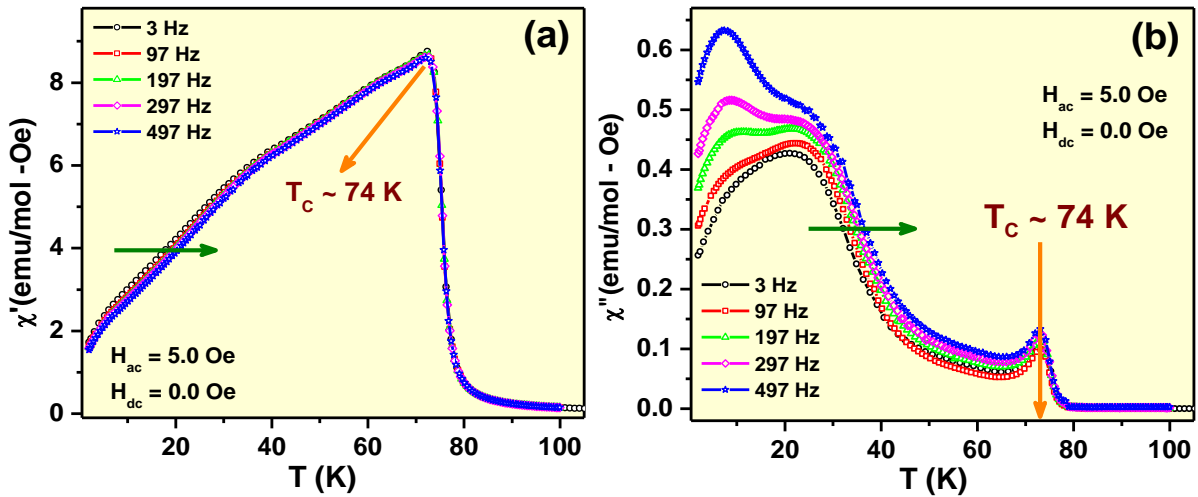

**Figure 3.** **(a)** Temperature dependence of in-phase susceptibility ($\chi'$), **(b)** and out of phase ac susceptibility ($\chi''$) of $Nd_3Al$ at 5 Oe ac field for various frequencies.

In the imaginary part of ACS, below $T_C$, the features completely change with frequency [fig. 3(b)]. $\chi''$ increases significantly with decrease in temperature below $T_C$, shows a broad maximum around 25 K, which shifts toward higher temperatures with frequency. Another peak is also observed at about 10 K, which shifts towards low temperatures with increase in frequency. Such features in χ" reflects energy losses associated with the magnetic domain walls/ the domain rotation in ferro/ferri magnets/ spin glass state/canted systems (23, 24). The variations of χ' and χ", even while corroborating the DC magnetization data, clearly show that the low temperature state of $Nd_3Al$ is far from a simple FM.

## Electrical resistivity and MR

As mentioned earlier, the most important observation from the $\rho - T$ plot is the high RRR value ($\rho_{300K}/\rho_{1.8K}$) of 83.2, which indicates the excellent purity/crystallinity of the sample, comparable to single crystals. The slope change in the zero field $\rho - T$ plot as shown in figure 4(a) corroborates with the $T_C$ derived from dχ/dT. Figure 4(a) also shows the shift in $T_C$ towards higher temperatures with field, along with a reduction of resistivity in the vicinity of $T_C$. Inset of fig. 4(a) shows the heating and cooling cycle data in zero field and the absence of thermal hysteresis denotes the absence of any structural (as in the case of $Ce_3Al$) or first order phase transition. Upon decreasing the temperature below 300 K, ρ(T) decreases similar to that in an itinerant system, but the temperature dependence is found to be non-linear with a concave curvature above 90 K. Such a curvature has been observed in the isostructural compounds $La_3Al$ (25) and $Pr_3Al$ (26). Other intermetallics like $Eu_3Ni_4Ga_4$ (27), RRhSn (28), RRhGe (29), $GdCu_6$ (30) and ternary half-Heusler compounds such as RPdSb (R= Gd and Tb) (31) also show concave curvature in resistivity. In such systems, the conduction electrons are scattered predominantly by the 5d band (Mott s – d scattering), due to the spin fluctuations and influence ρ(T) and the thermopower [S(T)] behaviour at high temperatures.

In presence of a field, it has also been found that, above 25 K all the resistivity curves show conventional FM nature. However, below 25 K anomalous features evolve. For a clear view, calculated MR vs. T is presented in figure 4(b), which shows the maximum negative MR ~ 15 % at $T_C$. Below 25 K, the MR is positive and has a value as high as 80 % at 13.5 T, which is rather unconventional for a FM. MR isotherms at different temperatures are shown in figure 4c. Another feature is the non-saturation of the positive MR with field, which is an indication of semi-metallic nature. By comparing the MR behaviour below 20 K with M-H and $\rho_{xy}$ curves (discussed below), we find that there is a strong correlation between the outcomes of these measurements. Furthermore, MR vs. H curves below and above 80 K have opposite curvatures, revealing a significant change in the nature of magnetic interactions across $T_C$ [Fig 4(c)].

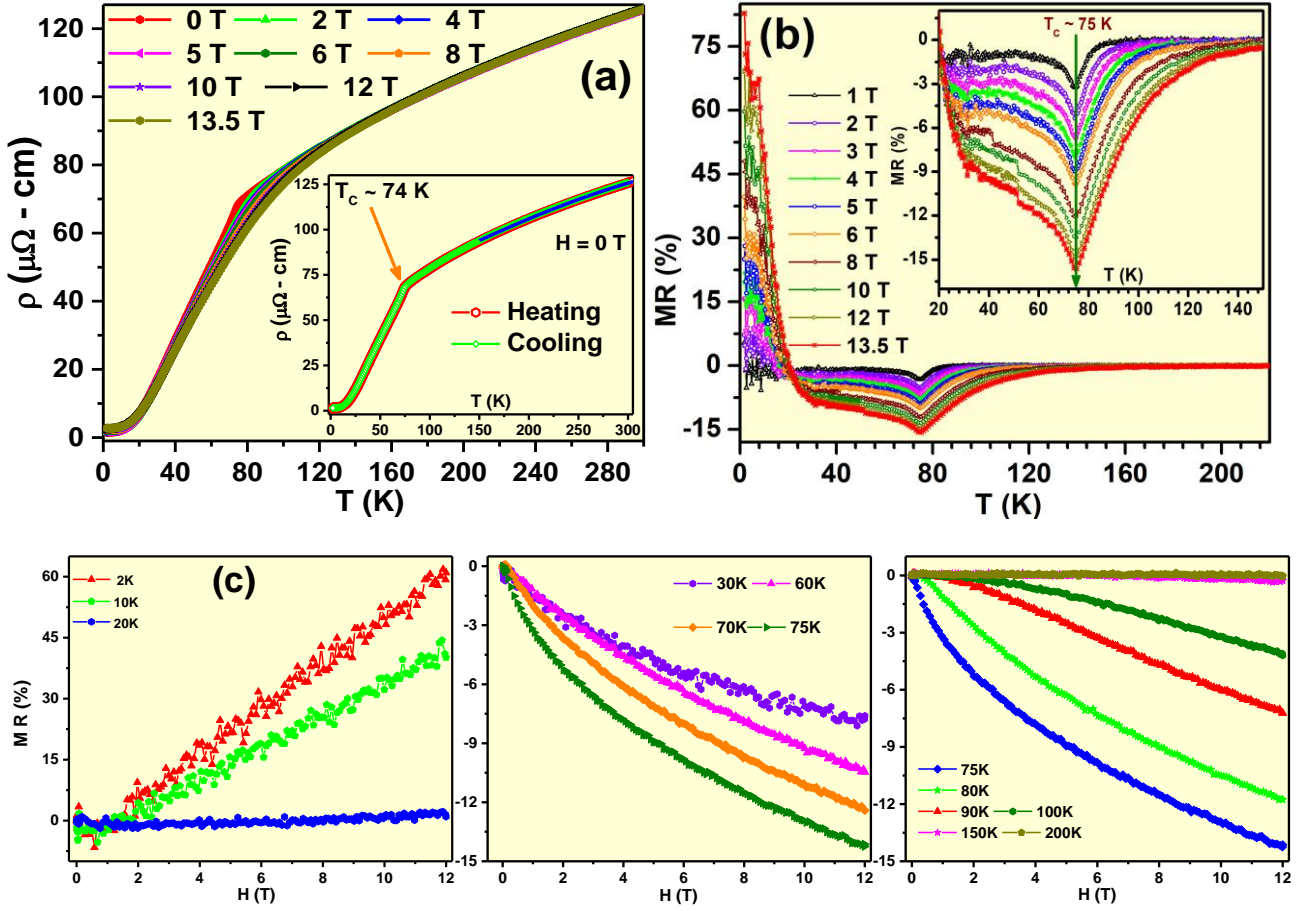

**Figure 4.** (a) ρ (T) vs. T up to 13.5 T that shows a slope change around 74 K. Inset: The absence of either structural or other first order transition in Nd$_3$Al is inferred from the non-hysteretic behavior. The solid line through the data points in inset show the fitting of the Bloch Gruneisen Mott relation to the experimental data. (b) MR vs. T at different fields. Inset shows MR near T$_C$. (c) MR vs. H at different temperatures.

To understand the origin of positive MR, the phonon contribution ($\rho_{ph}$) is extracted the from the ρ (T),. ρ - $\rho_{ph}$ is then fitted with equation (1) for ferromagnets with anisotropic magnon gap (Δ);

$$\Delta\rho = \rho - \rho_{ph} = \rho_0 + A.T^2 + D_m.\Delta.T\left(1 + 2\frac{T}{\Delta}\right)e^{-\left(\frac{\Delta}{T}\right)} \qquad (1)$$

where the first and second terms represent the residual and *e-e* interaction / spin fluctuation / spin glass contributions respectively (32). In the third term, $T^2$ signifies the FM contribution along with the magnon gap $\Delta$ while $D_m$ is the electron - magnon (spin disorder) scattering contribution. It also represents the spin-disordered scattering above T$_C$.

The increase in the residual resistivity $\rho_0$, with field suggests that the residual scattering is not simply due to potential scattering from impurities or lattice disorder, but involves the Nd - *4f* electrons i.e., spin fluctuation along with specific features of its electronic band structure. Here, we suspect that the increasing field enhances the frustration of the moments and hence drives the system to a partially disordered state, resulting in an enhanced spin disorder scattering due to the partially disordered (frustrated) moments. Hence, the values of the fitting parameters give a good indication about the field induced spin texture or partial frustration of spins. This gives rise to the monotonic enhancement of the parameter A (coefficient of $T^2$ term) and supports the anomalous magnetization as well as Hall resistivity (explained later) behaviour below 30 K.

**Anomalous and Topological Hall effects**

In fig. 5 (a) and (b) we show the Hall resistivity, $\rho_{xy}$ vs. H and Hall conductivity $\sigma_{xy}$ vs. H respectively, at different temperatures for Nd$_3$Al. Below 30 K, $\rho_{xy}$ curves vary in a non – linear fashion at low fields and linearly at high fields (i.e., above 2 T) with negative values. 30 K marks the boundary below which the sign of the Hall resistivity is negative and above which it is positive, irrespective of the field. Above 30 K, the data show the feature of an ordinary ferromagnet. This type of field dependence is also found in non – magnetic compounds like LaAl$_2$ and reflects the particular shape of the Fermi surface and the competing contributions of different electron-like and hole-like orbits (33). But in the present case, it is a FM compound in which the observed Hall effect arises from the Lorentz force (linear in the field) called ordinary Hall effect and the anomalous Hall effect arising from the intrinsic (due to the non-zero Berry curvature) and the extrinsic (skew scattering and side jumps associated with the spin-orbit coupling) contributions. A change of sign of Hall resistivity below 30 K indicates that the magnetic state here is no longer FM, but consists of some complex magnetic order. In addition, the nonlinear behavior as a function of field is quite evident below 30 K.

The most striking observation is that the anomalous Hall conductivity, $\sigma_{xy}$ (defined as $\frac{\rho_{xy}}{\rho_{xy}^2 + \rho_{xx}^2} \approx \frac{\rho_{xy}}{\rho_{xx}^2}$) below 30 K exhibits very high values, the highest being ~ 1.8 x 10$^5$ S/cm at 2 K/ 8 T. Such large values are seen only rarely and a few examples are kagome metal AV$_3$Sb$_5$ (A=K, Cs, Rb) (34) and Co$_3$Sn$_2$S$_2$ (8, 35). Above 30 K, the values are much smaller, i.e., ~ 600 S/cm, still larger compared to that of many FM metals and alloys. Shen et al. have reported a giant AHC contributed by both the intrinsic and the extrinsic contributions (36).

Non-linear behaviour in Hall data observed in Cu doped topological insulator Sb$_2$Te$_3$ i.e., Sb$_{1.90}$Cu$_{0.10}$Te$_3$ (37) has been attributed to the existence of AHE and the topological Hall effect (THE). A similar scenario appears to be present in Nd$_3$Al and hence we investigate the magnetic and transport behaviour from a topological angle, in detail, using electronic structure calculations as well.

To know the complexity in $\rho_{xy}$, it may be expressed as the combination of three terms:

$$\rho_{xy} = R_0 B + R_S M + \rho^{TH} \quad (2)$$

Here, $\rho_{xy}$ is the total Hall resistivity. $R_0$ and $R_s$ are the ordinary Hall effect (OHE) and anomalous Hall effect coefficients, respectively, and M is the magnetization. The third term, $\rho^{TH}$ represents the THE. Further, the coefficient of the AHE can be represented as:

$$R_S = a\rho_{xx} + b\rho_{xx}^2 \quad (3)$$

The linear and quadratic terms indicate the skew scattering and the side jump, respectively. Consistent with several reports

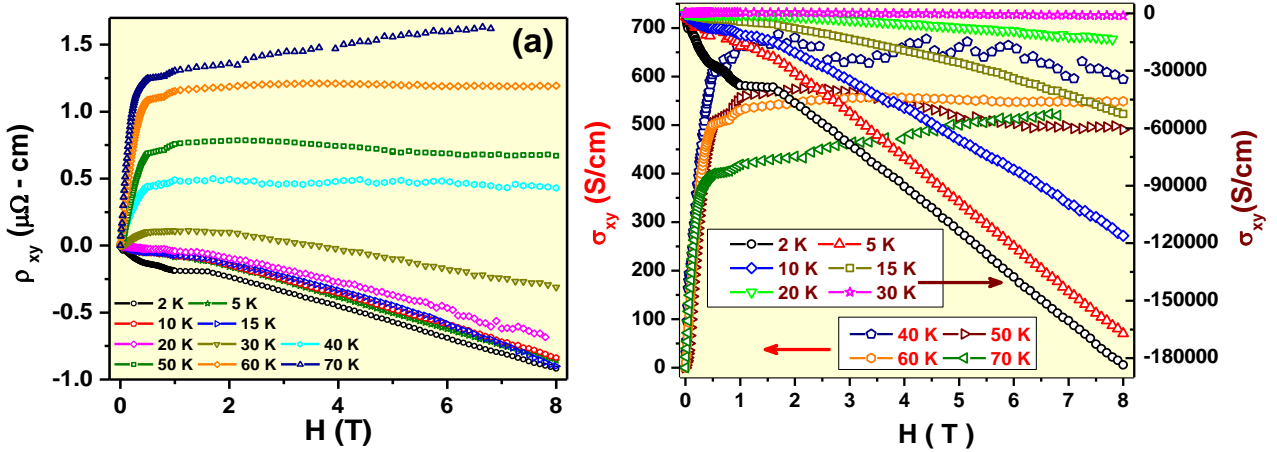

**Figure 5. (a)** Hall resistivity, $\rho_{xy}$ vs. H at different temperatures below 70 K. **(b)** Hall conductivity calculated as $\sigma_{xy} \sim \rho_{xy}/\rho_{xx}^2$. $\sigma_{xy}$ vs. H plots up to 30 K and above it, are shown on right and left scale respectively.

(38- 39), $R_S \propto \rho_{xx}^2$ and the linear term can be neglected (40 - 41). Also, $\rho^{TH}$ is absent at high magnetic fields and hence at high fields, equation (1) reduces to:

$$\rho_{xy} = R_0 B + b\rho_{xx}^2 M \quad (4)$$

By fitting the $\rho_{xy}$ vs. B graph at high magnetic fields, we can extract $R_0$ and b. It can be seen that the magnitude of the ordinary contribution is about two orders smaller than that of the sum of the other two contributions (AHE and THE). Below 30 K, we obtained a very good fit at high fields and obtained values of fitting parameters $R_0$ and b are shown in fig. 6 (a). Subtracting the ordinary and the anomalous Hall resistivity ($b\rho_{xx}^2 M$) contributions from the total Hall resistivity ($\rho_{xy}$) data in the full range of magnetic field, the topological Hall resistivity has been extracted. The signature of the existence of THE is observed at all temperatures below 30 K (Fig. 6b) and at 15 K / 0.02 T, it shows the maximum value of -0.375 $\mu\Omega - cm$. THE is negligibly small above 2 T for most of the temperatures except 15 K,

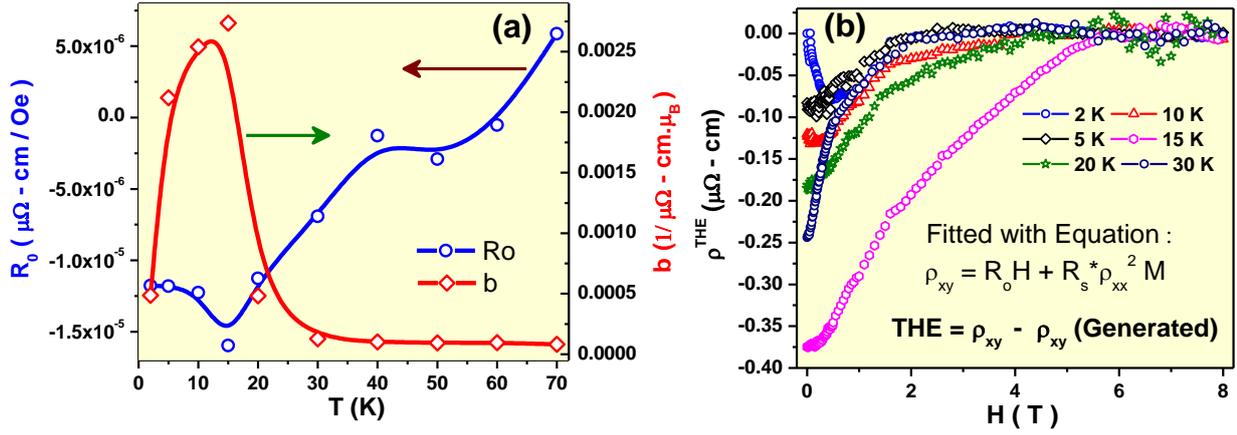

**Figure 6. (a)** Temperature dependence of parameters $R_0$ (left sale) and b (right scale) obtained from equation 4. **(b)** Topological Hall effect ($\rho^{THE}$) at different temperatures as a function of magnetic field.

As mentioned earlier, bifurcation in M vs. T and dispersion in ac χ vs. T curves below $T_C$ indicate a strong magnetocrystalline anisotropy or spin glass behavior below 30 K. From figure 7(a) it is clear that the enhancement in Hall conductivity and large positive magnetoresistance below 20 K are associated with reduction in magnetization. In a spin-glass system, the frustration due the exchange interactions leads to the noncollinear and noncoplanar spin configurations providing the spin-chirality mechanism of THE and the role of band (anti-) crossings near the Fermi energy can lead to a large intrinsic AHE. Such an observation has been reported in frustrated pyrochlore ferromagnet $Nd_2Mo_2O_7$ (42) and in $KV_3Sb_5$ (9). To know the role of spin - chirality or orientation for the origin of THE, we have plotted MR vs. $M^2$, as shown in Fig. 7 (b) which fits linearly. It clearly indicates that the origins of LPMR and THE are directly related to the abnormal spin texture in $Nd_3Al$.

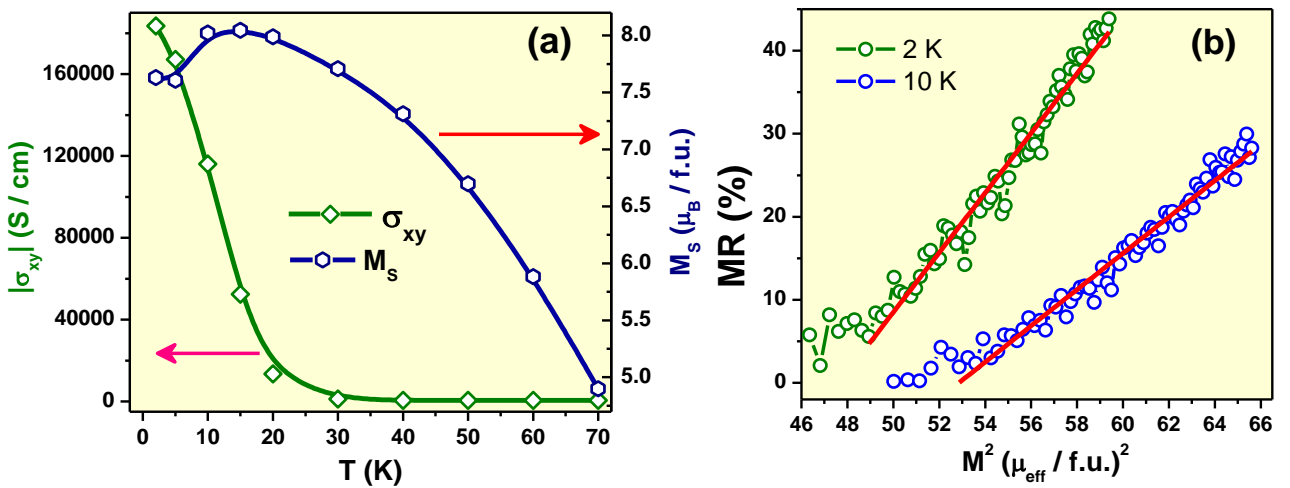

**Figure 7. (a)** Temperature dependence of $\sigma_{xy}$ on left scale and saturated magnetization on right scale at 8 T field. **(b)** Linear fit of MR vs $M^2$.

## Heat Capacity and Thermoelectric Power

Specific heat, $C_p(T)$ as a function of temperature in fields has been measured and shown in figure 8 (a) for a few representative fields. Analogously, as observed in M (T) and $\rho(T)$, a pronounced peak at 72 K (~ $T_C$) has also been observed, which got broadened with field with a shift towards higher temperature. The broadening of peak around the $T_C$ indicates the itinerant / short-range magnetic ordering or spin glass behaviour as supported by dc and ac $\chi$ data. In zero field, the estimated $\gamma$ and $\theta_D$ are $70 \pm 2$ mJ/molK$^2$ and 160 K ($\beta = 1.89$ mJ/mol-K$^4$) respectively. The calculated density of states (DOS), i.e., n($E_F$) near $E_F$ from $\gamma$, using the relation $n(E_F) = 3\gamma/(\pi k_B)^2$ is $29.71 \pm 0.85$ states/eV f.u., which is in very good agreement with the theoretically predicted density of states (n($E_F$)= 25 states/ eV f.u.). The high value of $\gamma$ in the present case as compared to other Nd and Al based alloys like Nd$_2$Al ($\gamma$ ~10 mJ mol$^{-1}$ K$^{-2}$) (43) and NdAl$_2$ ($\gamma$ ~ 9.5 mJ mol$^{-1}$ K$^{-2}$) (44), indicates that it belongs to a "false" heavy fermion system (45) and may be due to the spin fluctuation / spin glass behavior. Kadowaki-Wood's ratio (KWr), $A/\gamma^2$, which signifies the nature of electron correlations is usually applied for non-magnetic materials, but in view of the anomalous behavior of Nd$_3$Al with respect to magnetization and transport data, we have calculated KWr for Nd$_3$Al, which turns out to be 0.2 $\mu\Omega$cm.mol$^2$.K$^2$.J$^{-2}$ in zero field. This is found to be lower than that of certain transition metals and rare-earth based heavy fermion compounds (in the range of 0.4 -10 $\mu\Omega$-cm.mol$^2$.K$^2$.J$^{-2}$) (46). Such a low KWr hints at the non-enhancement of effective mass, ruling out the possibility of heavy electron formation in Nd$_3$Al. Therefore, the reason for the high $\gamma$ in Nd$_3$Al should be the itinerant nature of Nd - *4f* electrons, which is confirmed by the theoretical studies discussed later.

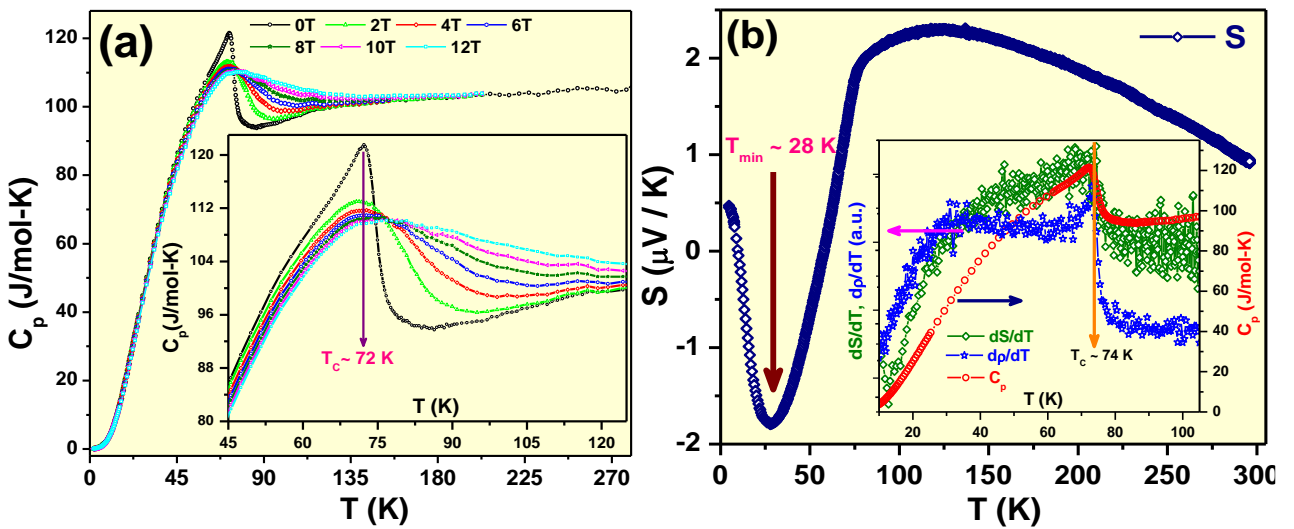

**Figure 8. (a)** $C_P$ vs. T of Nd$_3$Al in fields up to 12 T. Inset shows the specific heat of Nd$_3$Al around $T_C$. **(b)** S (T) vs. T of Nd$_3$Al. Inset shows the dS/dT (green) and d$\rho$/dT (blue) on left scale and $C_p$ (red) on right scale.

Being a good metal, the observed thermoelectric power shown in figure 8 (b) is quite contrasting. The positive sign of TEP in the paramagnetic phase indicates that the Fermi surface of this hexagonal alloy is not spherical but anisotropic in nature. This plays an important role in the TEP and the implications are rather complex in nature (47). The broad maximum around 125 K is attributed to a sharp peak in the DOS (common band features of $RE_3Al$ (48)) near the Fermi energy $E_F$. It has a sudden drop at $T_C$ and changes sign at 60 K, followed by an upturn below 28 K revealing the different types of exchange interactions between magnetic ions.

Furthermore, we have estimated the Sommerfeld-Wilson ratio, $R_W$:

$$R_W = \frac{\chi_o/3\mu_B^2}{\gamma_o/\pi^2 k_B^2}$$

(Using the value of $\chi_0 \approx 6.03 \times 10^{-3}$ emu/mole from the FCW dc magnetic susceptibility measured with 0.1 T field and $\gamma_0 = 70$ mJ/mole $K^2$) to be $\approx 79$. For a free electron gas system, $R_W = 1$. Thus, the Sommerfeld-Wilson ratio, which can give an estimate of the cluster moment (49) at low temperatures, is significantly enhanced for $Nd_3Al$. Kondo cluster glass system $CePd_{1-x}Rh_x$ for $0.8 \leq x \leq 0.87$ shows $R_W \sim 20-30$. (50) For $PrRhSn_3$ which has a very high $R_W \sim 1020$, has a strong electronic spin-spin interactions (For a free electron gas system, $R_W = 1$, hence strong spin correlations between itinerant Nd - $4f$ electrons result in $R_W \gg 1$) and strong FM fluctuations (51). An unusually high value of $R_W \sim 700$ is reported for cluster spin-glass system $(Sr_{1-x}Ca_x)_3Ru_2O_7$ for $x = 0.2$ (52). The nearly ferromagnetic systems that are known to exhibit the enhanced value of $R_W$ due to Stoner enhancement are $R_W = 40$ for $Ni_3Ga$ (53) and $R_W = 40$ for $Ca_{0.5}Sr_{0.5}RuO_4$ (54). An enhanced value of $R_W \sim 30$ is observed due to ferromagnetic quantum critical fluctuations in $YbRh_2(Si_{0.95}Ge_{0.05})_2$ (55). The observation of large $R_W$ for $Nd_3Al$ suggests that there may be significantly strong spin-spin interactions and FM fluctuations or reentrant spin glass behavior induced by disorder in the geometrically frustrated kagome lattice below 30 K (56).

Therefore, it is quite clear that, though $Nd_3Al$ is predominantly FM, its low temperature phase is rather complex. The reduction in the effective moment, thermomagnetic irreversibility, anomalous M-T and S-T dependence, positive sign of S, large and non-saturating positive MR, colossal AHC, existence of finite THE, anomalous values of KWr and $R_W$ ratios point towards topological band structure and topological spin texture in the frustrated kagome lattice. We also confirm the itinerant FM nature of this compound, but rule out the possibility of heavy fermion or pure simple spin glass state. In order to throw some more light on the topological nature, we have carried out band structure calculations as described below.

## Band structure calculations

Figure 9(a) shows the 2D projected crystal structure of Nd$_3$Al. As mentioned earlier, it forms a kagome unit cell coordinated by Nd atoms, where Al atoms occupy the centers of the hexagons. Figure 9(b) shows the bulk Brillouin Zone. Table I shows the theoretically optimized lattice parameters, formation energy, atom projected and total magnetic moment of Nd$_3$Al. These lattice parameters and the total magnetic moment match fairly well with those of experiment. Figure 10 shows the electronic band structure and DOS obtained after including the effect of spin orbit coupling. The electronic structure clearly confirms that Nd$_3$Al is a FM metal.

Nd$_3$Al structure possesses a triangular geometry involving Nd-atoms. A close inspection of the band structure at/near the Fermi level (see Fig. 9(c)) clearly indicates the appearance of Weyl points (WPs) (in pair) with opposite chirality off the high symmetry lines. The precise positions of the WPs have been obtained by calculating the Wilson-loop evolution (57). Apart from this, there are also several flat bands near E$_F$ which are responsible for the peak in the DOS. The major contribution to both the flat bands as well as the Weyl nodes arises from the itinerant 4*f*-electrons of Nd in agreement with the experimental observations. Such peaks in the DOS at/near E$_F$ are expected to be responsible for the enhanced Sommerfeld coefficient as revealed by the experimental heat capacity data. However, because these WPs [in Fig. 10(c)] are mixed with other bulk states, it is not easy to clarify the details of how the Fermi arcs will connect the WPs. The existence of WPs near E$_F$ can give rise to very high carrier mobilities and hence responsible for large AHE and other unusual transport behavior at low temperatures in this system (8, 10 – 11).

As mentioned in the earlier sections, our most significant experimental observation is the colossal anomalous Hall conductivity at low temperatures, that too in a polycrystalline sample. In general, large AHC can arise from a variety of effects. One particular interesting limit is when the anomalous Hall angle approaches 90°; a characteristic of the intrinsic quantum AHE observed in time reversal symmetry breaking topological insulators (58- 60). The intrinsic AHE is controlled by the electronic structure of a material, which causes an electron to possess a transverse momentum as it travels in between scattering events (61). This mechanism (arising from the occurrence of WPs) plays a dominant role in various topological materials where AHC is of the order of 1000 $\Omega^{-1}$cm$^{-1}$(8, 62). On the other hand, the extrinsic AHE originates from the electron scattering involving sudden changes in the periodic potential of a crystal, caused by structural defects or chemical/magnetic impurities. Apart from this, local groups of coupled spins, tilted magnetic clusters are also known to contribute in this effect. Magnetic ions in a kagome net are also reported to generate enhanced skew scattering and thus a large AHE (63). Crystals with triangular lattice having geometrically frustrated magnetism and spin liquids are particularly likely to exhibit this type of AHC. A plausible way to realize such a mechanism is to introduce electronic topology into magnetically frustrated systems. This is fundamentally different as compared to the Berry phase mechanism, which is related to the band topology.

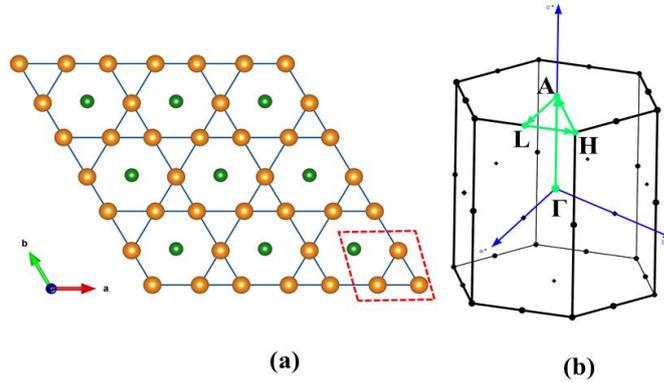

(a) (b)

**Figure 9.** For $Nd_3Al$ **(a)** The kagome layer involving Nd (orange spheres) and Al atoms (green spheres). The red dashed rhomboid with the corner-sharing triangles demonstrates the unit cell of the kagome lattice. **(b)** Schematic bulk Brillouin zone, where green arrows depict the selected high symmetry k-path direction used to study the bulk band structure.

In addition, our kagome system $Nd_3Al$ appears to have large contributions from the topological spin textures, which in turn contributes to the AHC and the topological component, as observed experimentally. A somewhat similar scenario has been seen in systems like MnGe films in which the longitudinal conductivity is quite high and the AHC is of the order of $4\times10^4$ S/cm (64).

TABLE I. For $Nd_3Al$, theoretically optimized lattice constants (in Angstrom), total and atom-projected magnetic moments (in $\mu_B$) and formation energy ($\Delta E_f$).

| $m^{Nd}$ | $m^{Al}$ | $m^{Total}$ | Lattice parameters (Å) | $\Delta E_f$ (eV/f.u.) |
|---|---|---|---|---|
| 3.4 | 0.00 | 10.2 | a=b=7.1, c=5.44 | -0.52 |

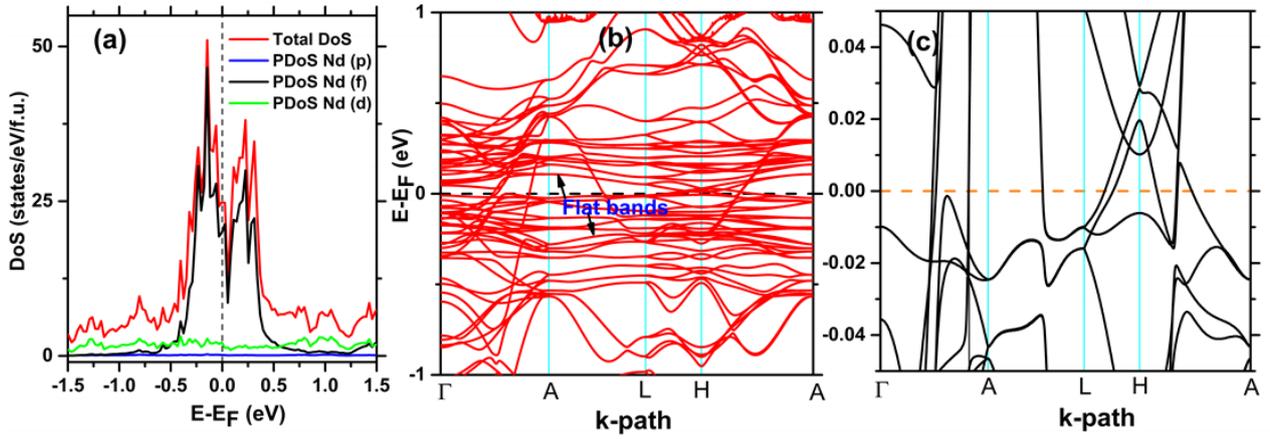

**Figure 10** PBE+SOC band structure and DOS for Nd$_3$Al. **(a)** Total and atom/orbital projected DOS. **(b)** Electronic band structure along the high symmetry k-path Γ-A-L-H-A. **(c)** A zoomed-in view of **(b)** near the Fermi level (F$_F$). There are two pairs of Weyl points aroun the H-point with opposite chirality.

## Conclusions

In summary, a colossal anomalous Hall effect ($\sim 1.8 \times 10^5$ S/cm) and a large positive magnetoresistance (~80%) along with topological Hall effect is observed in a highly conductive ferromagnetic metal candidate, Nd$_3$Al. To the best of our knowledge, this is the highest anomalous Hall conductivity reported ever. Nd$_3$Al structure showcase a beautiful kagome net formed by Nd atoms. M vs. H and MR vs. H isotherms at different temperatures reveal an anomalous behavior below the T$_C$. Our detailed experimental measurements along with the band structure calculations give strong indication to the itinerant nature of the Nd - *4f* electrons. Topological non-trivial behavior of Nd$_3$Al is revealed by the semimetallic nature of the MR data as well the occurrence of Weyl points in the calculated band structure. Along with the intrinsic Berry phase contribution that is known to be quite large in kagome systems, a spin-orbit field generated via the electronic topology and enhanced skew scattering due to the kagome lattice are also proposed to be the reason for such a colossal anomalous Hall conductivity. Furthermore, it is seen to acquire an abnormal spin texture at low temperatures and in moderate fields, which gives rise to THE. In short, we believe that such a combination of exotic band structure and a skyrmionics-like spin texture in a geometrically frustrated metallic system provides a novel platform to study topological matter. Another advantage of such a system is the possibility of further enhancing the AHE by modifying the skew scattering potential in situ. This is possible because magnetic fluctuations are tunable via the external perturbations. Though we have presented many experimental evidences towards the complex magnetic state as well as the special topological features in Nd$_3$Al, synthesis of single crystals and the use of experimental probes such as ARPES are planned for future exploration.

# EXPERIMENTAL AND COMPUTATIONAL DETAILS

Polycrystalline samples of $Nd_3Al$ and its nonmagnetic analogue $La_3Al$ (for calculating the non-magnetic contribution to the heat capacity) were prepared by arc melting technique. The constituent elements of purity at least 99.9% were melted in an argon atmosphere using titanium ball as a getter. The phase purity of the sample was confirmed by powder x-ray diffraction (XRD). Resistivity ($\rho$), Hall effect, heat capacity ($C_p$) and DC magnetization (M) have been measured using QD PPMS with VSM option. AC susceptibility measurements were performed using the SQUID magnetometer in the temperature range of 2–120 K and in the frequency interval of 3 - 500 Hz. Thermoelectric power (TEP) in zero field was measured using differential dc sandwich method in a homemade set up (4–300 K) (65).

Ab-initio calculations were performed using the spin resolved density functional theory (DFT) (66) as implemented within Vienna ab initio simulation package (VASP) (67-69) with a projected augmented-wave (PAW) basis (70). Pseudopotential formalism with Perdew, Burke and Ernzerhof (PBE) exchange-correlation functional (71) was used for the electronic structure calculations. To perform the Brillouin zone (BZ) integration, a $9 \times 9 \times 11$ $\Gamma$-centred k-mesh within the tetrahedron method was used (72). A plane wave energy cut-off of 500 eV was used for all the calculations. All the structures were fully relaxed with the total energies (forces) converged to values less than $10^{-6}$ eV (0.01 eV/Å). The theoretically optimized lattice parameters are $a = b = 7.1$ Å, and $c = 5.44$ Å, which are in fair agreement with our experimental results. The effect of spin-orbit coupling (SOC) has been included in all the calculations.

## Acknowledgements

The authors thank UGC DAE CSR Indore for electrical and thermal transport measurements. Dr. Rajeev Rawat is thanked for his valuable suggestions. DS and JN thank IIT Bombay for providing the financial support. KGS thanks SERB for the financial support via a sponsored project with grant number CRG/2020/005589. Authors thank Dr. C K Barman for helping set up band structure calculations.